\documentclass{moriond}

\usepackage{amsfonts,amsmath,slashed,mathtools,color}

\newcommand{\beq}{\begin{equation}}
\newcommand{\eeq}{\end{equation}}

\newcommand{\Photo}{}

\begin{document}
\vspace*{4cm}
\title{Connecting neutral current $B$ anomalies with the heaviness of the third family}

\author{Joe Davighi}

\address{DAMTP, University of Cambridge, United Kingdom}

\maketitle
\abstracts{It is possible, and for several reasons attractive, to explain a collection of recent anomalies involving $b\rightarrow s\mu\mu$ processes with a $Z^{\prime}$ gauge boson coupled only to the third family in the weak eigenbasis. From this premise, requiring cancellation of all gauge anomalies (including mixed and gravitational anomalies) fixes a unique charge assignment for the third family Standard Model fermions, which is simply proportional to hypercharge. After a brief discussion of some general features of anomaly cancellation in $Z^\prime$ theories, we discuss the phenomenology of such a `Third Family Hypercharge Model', which is subject to a trio of important constraints: (i) $B_s-\bar{B}_s$ mixing, (ii) lepton universality of the $Z$ boson couplings, and (iii) constraints from direct searches for the $Z^\prime$ boson at the LHC. Finally, in gauging third family hypercharge, this model forbids all Yukawa couplings  (at the renormalisable level) save those of the third family, leading to a possible explanation of the heaviness of the third family.}

\section{Introduction}

There is a tension between Standard Model (SM) predictions and experimental measurements involving $b\rightarrow s\mu\mu$ transitions, for example in LHCb's measurements of the lepton flavour universality (LFU) ratios $R_{K^{(\ast)}}=BR(B\rightarrow K^{(\ast)} \mu^+\mu^-)/BR(B\rightarrow K^{(\ast)} e^+ e^-)$. For the di-lepton invariant mass-squared bin $q^2\in \left[1.1,6\right]\mathrm{GeV}^2$, in which the SM predicts $R_{K^{(\ast)}}$ equal to unity at the percent level, the new measurement of $R_K$ (which includes Run-1 data and 2 fb$^{-1}$ of Run-2 data) is $R_K=0.846^{+0.060+0.016}_{-0.054-0.014}$, where the first (second) uncertainty is statistical (systematic).\cite{LHCbMoriond} LHCb has also measured  $R_ {K^\ast}=0.69^{+0.11}_{-0.07}\pm 0.05$ in the same $q^2$ bin, and $R_ {K^\ast}=0.66^{+0.11}_{-0.07}\pm 0.03$ for $q^2\in \left[0.045,1.1\right]\mathrm{GeV}^2$, both using Run-1 data only.\cite{Aaij:2017vbb} 
Furthermore, there are notable deviations between the SM prediction and the measurements of $BR(B_s\rightarrow \mu\mu)$ by LHCb \& CMS,\cite{CMS:2014xfa} and (as of Moriond 2019) ATLAS,\cite{Aaboud:2018mst} and in $B\rightarrow K^\ast \mu^+\mu^-$ angular observables such as $P_5^\prime$.\cite{Aaij:2015oid,Khachatryan:2015isa} This collection of discrepancies, which we shall henceforth refer to as the `neutral current $B$-anomalies' (NCBAs), all point consistently towards a common new physics explanation in which LFU is violated, favouring (for example) a reduction in the effective coupling of the left-handed $b\bar{s}$ current to muons.\cite{Aebischer:2019mlg} The absence of similar anomalies in semileptonic decays of lighter mesons, such as kaons, pions, or charm-mesons, hints that whatever new physics underlies the NCBAs couples primarily to the third-family quarks. Taking this hint seriously, we here outline a simple model in which a heavy $Z^\prime$ boson is coupled to the third family.

\section{Third family $Z^\prime$ models}

Let us suppose that the NCBAs are mediated by a heavy $Z^\prime$ boson, deriving from a spontaneously broken $U(1)^\prime$ gauge symmetry by which we extend the SM, under which only the third family will be charged in the weak eigenbasis. In addition to the $Z^\prime$, we require a scalar which is charged only under $U(1)^\prime$, responsible for breaking $U(1)^\prime$ at the TeV scale. In the spirit of bottom-up model building, we shall not introduce any further fields beyond those of the SM. We nonetheless need additional input to constrain the third family charges. 

When building such a low-energy effective field theory (EFT), it is prudent to insist on gauge anomaly cancellation. This avoids the complication of including appropriate Wess-Zumino (WZ) terms to cancel anomalies in an otherwise anomalous low-energy EFT. Moreover, even if a specific set of anomalies can be cancelled by new UV physics, such as a set of heavy chiral fermions, it will be difficult to give these chiral fermions heavy enough masses in a consistent framework.
Thus, we shall require that our charge assignment is anomaly-free. 

\subsection{Anomaly cancellation in SM$\times U(1)^\prime$ theories}

The space of anomaly-free SM$\times U(1)^\prime$ theories has been explored in detail recently.\cite{Allanach:2018vjg} The interest in such flavoured $Z^\prime$ models goes beyond the NCBAs; for example, in modelling dark matter\cite{Altmannshofer:2016jzy} or fermion masses.\cite{Froggatt:1978nt} Thus, before we define our third family $Z^\prime$ model, we shall briefly discuss anomaly cancellation in $Z^\prime$ model building more generally.

Given only the chiral fermions of the SM, there are fifteen rational $U(1)^\prime$ charges $F_{X_i}$ to assign in a SM$\times U(1)^\prime$ theory, where $X\in\{Q,L,e,u,d\}$, and $i\in\{1,2,3\}$ labels the family. After rescaling the gauge coupling, anomaly cancellation implies a system of Diophantine equations over fifteen integer variables. These equations are non-linear, due to the $U(1)_Y\times U(1)^{\prime 2}$ and $U(1)^{\prime 3}$ anomalies, which makes parametrizing its solution space a challenging arithmetic problem. Nonetheless, if we restrict to just two families of the SM, we can parametrize all solutions analytically using Diophantine methods.\cite{Allanach:2018vjg} To wit, the sums of charges $F_{X+}\equiv F_{X_1}+F_{X_2}$ must be proportional to hypercharge, \emph{viz.} $F_{u+}=4F_{Q+}$, $F_{d+}=-2F_{Q+}$, $F_{e+}=-6F_{Q+}$, and $F_{L+}=-3F_{Q+}$, where $F_{Q+} \in \mathbb{Z}$, and the differences of charges $F_{X-}\equiv F_{X_1}-F_{X_2}$ are fixed by the quadratic equation $F_{Q-}^2 + F_{d-}^2 + F_{e-}^{2} - F_{L-}^2 - 2 F_{u-}^2=0$. 
All integer solutions to this equation are parametrized by four positive integers $\{ a, a_e, a_d, a_u \}$, explicitly 
$F_{Q-}=a^2-a_d^2-a_e^2+2a_u^2$, $F_{L-}=a^2+a_d^2+a_e^2-2 a_u^2$, $F_{d-}=2aa_d$, $F_{e-}=2aa_e$, and $F_{u-}=2aa_u$.

In the full three-family SM, 
we find a vast `atlas' of anomaly-free SM$\times U(1)^\prime$ theories,\cite{Allanach:2018vjg} some small fraction of which have been explored in the literature,\footnote{For example, gauging $L_\mu - L_\tau$ has been extensively explored phenomenologically.\cite{Altmannshofer:2015mqa}} by finding all solutions with integer charges of magnitude up to some pre-defined maximum $Q_{\mathrm{max}}$ using a numerical scan. We find that, for example with $Q_{\mathrm{max}}=6$, there are more than $10^5$ inequivalent (up to rescalings and permuting families) charge assignments if three right-handed neutrinos are included. On the other side of the coin, anomaly cancellation is a stringent constraint on $U(1)^\prime$ charges; with $Q_{\mathrm{max}}=6$, including right-handed neutrinos, only about one in every billion possible charge assignments happens to be anomaly-free.
This unexplored solution space only opens up in the full three-family SM. We now return to the special case with only the third family charged under $U(1)^\prime$. It turns out that in this case, there is a unique anomaly-free charge assignment, and that is simply hypercharge.

\subsection{Third Family Hypercharge Model and the heaviness of the third family}

In the Third Family Hypercharge Model (TFHM),\cite{Allanach:2018lvl} the charges of the third family fields in the weak eigenbasis equal their hypercharges, with the first two families being uncharged under $U(1)^\prime$. If we assign the Higgs a $U(1)^\prime$ charge also equal to its hypercharge, then the only gauge invariant Yukawa couplings are those of the third family. In the spirit of EFT, we nonetheless expect a perturbation around this renormalizable Yukawa sector due to higher-dimension operators. While an explanation of the precise hierarchies observed in the quark and lepton masses, and in the mixing angles of the CKM and PMNS matrices, would require more detailed model building of the UV physics, the zeroth-order predictions of such a setup are that (i) the third family is hierarchically heavier than the first two, and (ii) quark mixing angles are small,\footnote{Note that lepton mixing is not expected to be small, because we have not specified a mass sector for neutrinos.} thus shedding light on the coarsest features of the SM flavour problem.

In order to compare the model with experimental bounds, one must specify the mixing between the mass and weak eigenbases. Consider the limiting case defined by the mixing matrices:
\beq
V_{d_L}=\begin{pmatrix}
1 & 0 & 0 \\
0 & \cos \theta_{sb} & -\sin \theta_{sb} \\
0 & \sin \theta_{sb} & \cos \theta_{sb} \\
\end{pmatrix},
\qquad
V_{e_L}=\begin{pmatrix}
1 & 0 & 0 \\
0 & 0 & 1 \\
0 & 1 & 0 \\
\end{pmatrix},
\eeq
together with $V_{u_L}=V_{d_L}V^\dagger$, $V_{u_R}=V_{d_R}=1$,  $V_{\nu_L}=V_{e_L}U^\dagger$, and $V_{e_R}=1$, where $V$ is the CKM matrix and $U$ is the PMNS matrix.\footnote{Note that this choice of $V_{e_L}$ implies that the tauon Yukawa must in fact be suppressed relative to the na\"ive order one expectation. We will address this issue in a future work.} We shall refer to this particular one-parameter ($\theta_{sb}$) family of example cases of the Third Family Hypercharge Model as the `TFHMeg'.

\section{Phenomenology of the TFHM example case}

\subsection{Constraints}

\begin{figure}
\begin{center}
\unitlength=15cm
\begin{picture}(0.8,0.35)
    \put(-0.25,-0.03){\includegraphics[width=0.7\textwidth]{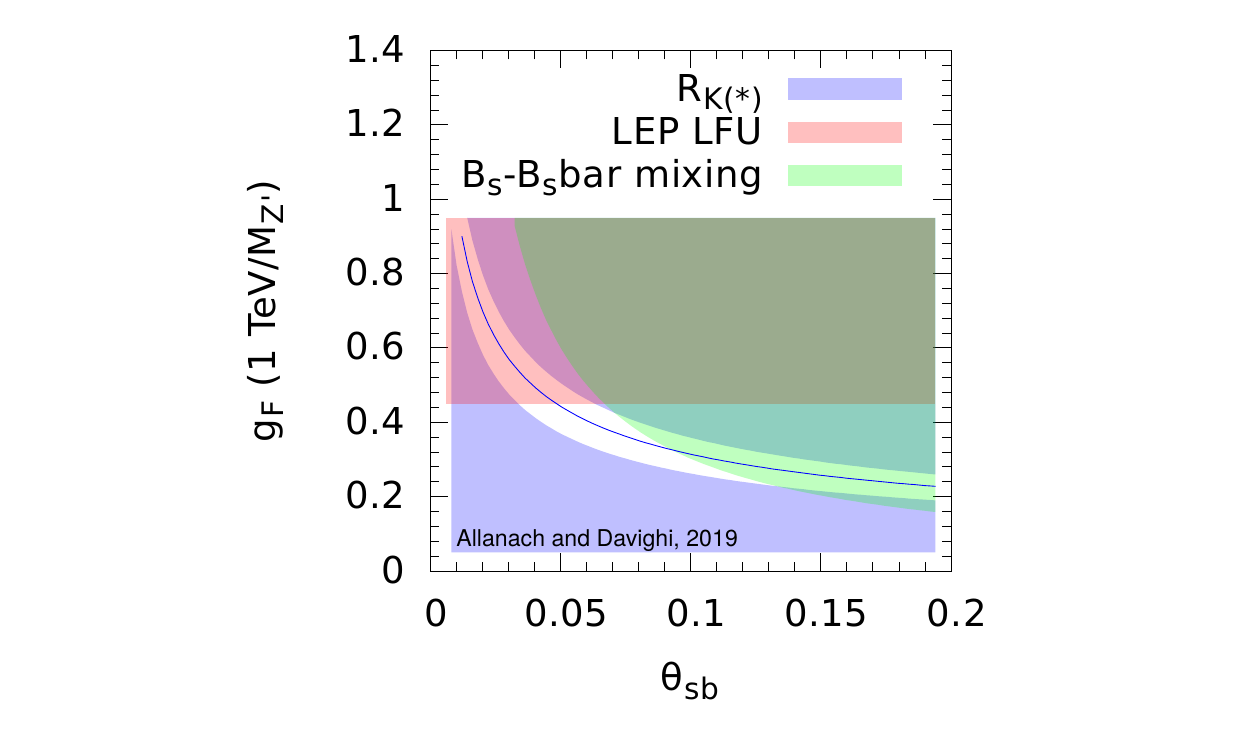}}
   \put(0.30,-0.03){\includegraphics[width=0.7\textwidth]{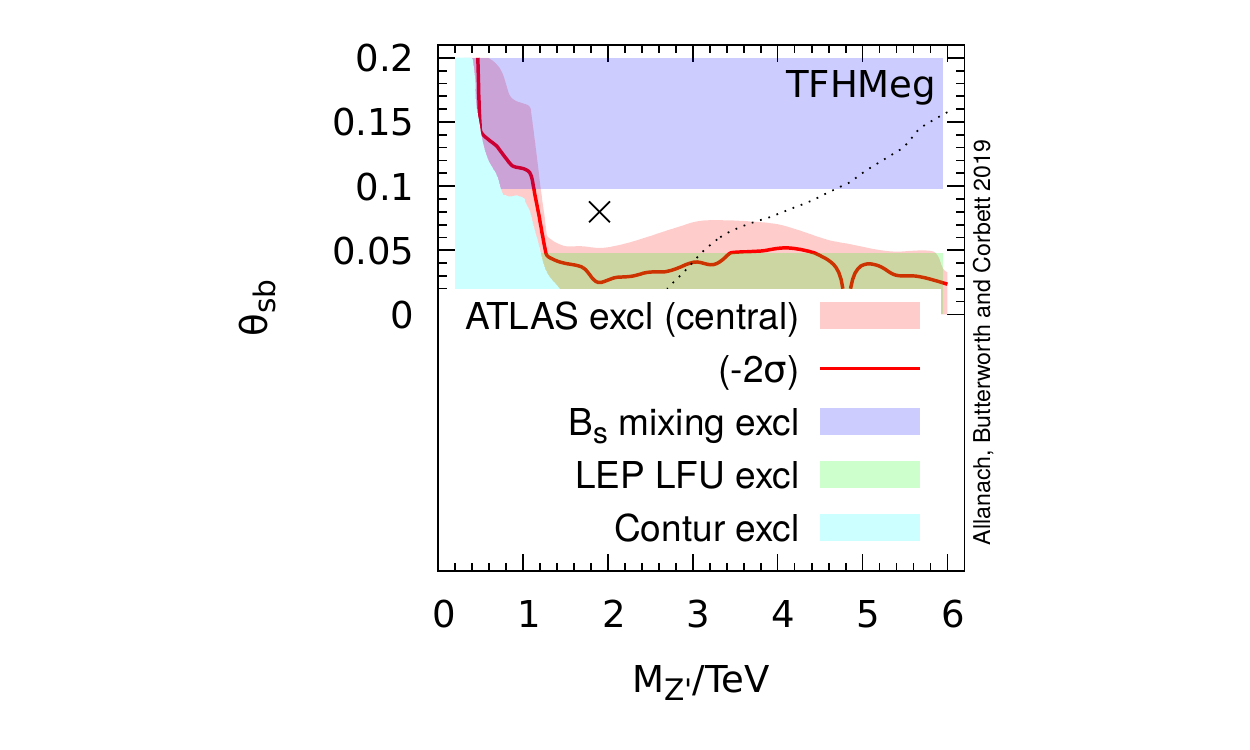}}
\end{picture}
\end{center}
\caption{Bounds on the TFHMeg; in both plots, the white region is allowed parameter space. Left - the bounds on $g_F/M_{Z^\prime}$ versus $\theta_{sb}$ from fitting the (post Moriond 2019) NCBAs (blue), including constraints from LEP LFU (red) and $B_s-\overline{B_s}$ mixing (green). Right - we also include the constraint from direct $Z^\prime \rightarrow \mu\mu$ searches at the ATLAS, in the $\theta_{sb}$ \emph{vs.} $M_{Z^\prime}$ plane.\protect\cite{Allanach:2019mfl} Here, the value of the coupling $g_F$ is fixed to the central value from the fit to the NCBAs. Also, we include constraints from other LHC searches using the \texttt{CONTUR} tool (turquoise).\protect\cite{Allanach:2019mfl,Butterworth:2016sqg}}
\label{TFHM}
\end{figure}

The TFHMeg lagrangian contains the operators $g_{sb}\overline{s} \slashed{Z}^\prime P_L{b}+g_{\mu\mu}\bar{\mu} \slashed{Z}^\prime P_L \mu + H.c$, where $g_{sb}=g_F (\sin 2 \theta_{sb})/12$ and $g_{\mu\mu}=-g_F/2$, which, after integrating out the $Z^\prime$, generate effective operators that can provide a good fit to the NCBAs.
Using the post-Moriond 2019 global fit to the NCBA data,\cite{Aebischer:2019mlg}
the bound on the TFHMeg is $g_F=\left(M_{Z^\prime}/\mathrm{36~TeV}\right) \sqrt{24x/\sin (2\theta_{sb})}$, where $x=1.06 \pm 0.16$, at the 95\% CL.\cite{Allanach:2019mfl}

The other important constraints on this model are threefold. Firstly, the $g_{sb}$ coupling of the $Z^\prime$ leads to a tree-level contribution to $B_s-\overline{B_s}$ mixing, which is loop-suppressed in the SM. While there are a number of different calculations, the most recent constraint, which incorporates lattice data and sum rules\cite{King:2019lal} with experimental measurements,\cite{Amhis:2016xyh} yields the bound $|g_{sb}| \leq M_{Z^\prime}/(194\mathrm{~TeV})$.\cite{Allanach:2019mfl}
Secondly, in this model there is mass-mixing between the $Z$ and $Z^\prime$, because the Higgs has $U(1)^\prime$ charge. While this mixing is small,
the resulting flavour-dependent couplings inherited by the $Z$ boson are tightly constrained by LEP data. In particular, the LEP measurement of $R\equiv \Gamma(Z\rightarrow e^+ e^-)/\Gamma(Z\rightarrow \mu^+ \mu-)$ results in the bound $g_F<M_{Z^\prime}/(2.2\mathrm{~TeV})$, at the 95\% CL.\cite{Allanach:2018lvl} Finally, there is a constraint coming from direct searches for the $Z^\prime$ at colliders, for example in the dimuon decay channel. This constraint is obtained by recasting the most recent $Z^\prime\rightarrow \mu^+ \mu^-$ search constraints from ATLAS,\cite{Aad:2019fac} which uses 139 fb$^{-1}$ of 13 TeV $pp$ collisions at the LHC.\cite{Allanach:2019mfl}
These constraints leave a viable region of parameter space in the TFHMeg (Fig.~\ref{TFHM}).

\subsection{Predictions}

In addition to direct $Z^\prime\rightarrow \mu^+ \mu^-$ searches, there are other distinct predictions of the TFHMeg (and the TFHM in general). Firstly, the $Z^\prime$ decays predominantly to third family fermions, with the largest branching ratios to $t\bar{t}$ (42\%) and $\tau^+ \tau^-$ (30\%). Nevertheless, the bounds from dimuon searches (branching ratio of 8\%) provide the strongest constraint at present.\cite{Allanach:2019mfl} With the nominal integrated luminosity expected at the HL-LHC being 3000 fb$^{-1}$, we expect the parameter space of the TFHMeg to be fully covered by the HL-LHC.\cite{Allanach:2019mfl}
In addition to these exciting prospects from direct searches at the LHC, the TFHM also predicts rare top decays, $t\rightarrow Zu$ and $t\rightarrow Zc$, as a result of flavour-changing $Z^\prime$ couplings to up-type quarks and the $Z-Z^\prime$ mixing.\cite{Allanach:2018lvl} The current constraints from LHC bounds on $BR(t\rightarrow u,c)$ are weak, but likely to become important in the HL-LHC. Finally, the TFHMeg predicts a deficit in $BR(B\rightarrow K^{(\ast)} \tau^+\tau^-)$.\cite{Allanach:2018lvl} Advances in $\tau$ identification and measurements of, for example, the LFU-probing ratio 
$BR(B\rightarrow K \tau^+\tau^-)/BR(B\rightarrow K e^+e^-)$ are much anticipated at both LHCb and Belle II.

\section{Conclusion}

We have discussed the possibility that the NCBAs might be mediated by a $Z^\prime$ boson coupled only to the third family in the weak eigenbasis, with charges set to hypercharge by anomaly cancellation. We saw how gauging third family hypercharge might explain the heaviness of the third family. Finally, such a model leads to a distinctive and testable phenomenology, with resonances predicted in top, bottom, and tauon pairs.

\section*{Acknowledgments}

JD thanks his collaborators Ben Allanach and Scott Melville for their contributions to this work, and for their permission to present the figures used in this talk. JD is supported by The Cambridge Trust and by the STFC consolidated grant ST/P000681/1.

\end{document}